%% file: main.tex
\documentclass[sigconf]{acmart}
\renewcommand\footnotetextcopyrightpermission[1]{}

\include{__packages__}

\AtBeginDocument{%
  }

\setcopyright{acmlicensed}
\copyrightyear{2024}
\acmYear{2024}
\acmDOI{XXXXXXX.XXXXXXX}

\acmConference[MODEVAR'24]{6th International Workshop on Languages for Modelling Variability (MODEVAR 2024)}{February 6, 2024}{Bern, Switzerland}
\acmISBN{978-1-4503-XXXX-X/18/06}

\begin{document}

\title{UVL Sentinel: a tool for parsing and syntactic correction of UVL datasets}

\author{David Romero-Organvidez}
\affiliation{%
    \institution{Dept. of Computer Languages and Systems}
    \institution{University of Seville}
    \streetaddress{Av. Reina Mercedes S/N}
    \city{Seville}
    \country{Spain}
}
\email{drorganvidez@us.es}

\author{Jose A. Galindo}
\affiliation{%
    \institution{Dept. of Computer Languages and Systems}
    \institution{University of Seville}
    \city{Sevilla}
    \country{Spain}
}
\email{jagalindo@us.es}

\author{David Benavides}
\affiliation{%
    \institution{Dept. of Computer Languages and Systems}
    \institution{University of Seville}
    \streetaddress{Av. Reina Mercedes S/N}
    \city{Seville}
    \country{Spain}
}
\email{benavides@us.es}

\renewcommand{\shortauthors}{David Romero-Organvidez et al.}

\begin{abstract}


Feature models have become a \emph{de facto} standard for representing variability in software product lines. UVL (Universal Variability Language) is a language which expresses the features, dependencies, and constraints between them. This language is written in plain text and follows a syntactic structure that needs to be processed by a parser. This parser is software with specific syntactic rules that the language must comply with to be processed correctly. Researchers have datasets with numerous feature models. The language description form of these feature models is tied to a version of the parser language. When the parser is updated to support new features or correct previous ones, these feature models are often no longer compatible, generating incompatibilities and inconsistency within the dataset. In this paper, we present UVL Sentinel. This tool analyzes a dataset of feature models in UVL format, generating error analysis reports, describing those errors and, eventually, a syntactic processing that applies the most common solutions. This tool can detect the incompatibilities of the feature models of a dataset when the parser is updated and tries to correct the most common syntactic errors, facilitating the management of the dataset and the adaptation of their models to the new version of the parser. Our tool was evaluated using a dataset of 1,479 UVL models from different sources and helped semi-automatically fix 185 warnings and syntax errors. 

\end{abstract}



\maketitle

\section{Introduction}

UVL (Universal Variability Language) is a community proposal to establish a universal textual language for feature modelling~\cite{uvltutorial}. This language describes feature models and facilitates the exchange of scientific knowledge. It is also the basis for feature model analysis tools such as FeatureIDE~\cite{Sundermann2021_UVLFeatureIDE} or Flama~\cite{galindo_flama_splc_23}. 

There are numerous datasets containing an arbitrary number of feature models. These feature models are extracted from various sources and transformed into the UVL language. The feature models in UVL are processed by a parser, a software that parses the model description. The parser validates that specific defined syntactic rules are met and reports possible errors in the model syntax. 

Listing \ref{lst:uvl_example} shows a UVL file representing a mobile phone feature model. This feature model represents the constraints and dependencies between features, such as \emph{Camera}, \emph{Audio\_Formats} and \emph{Camera\_Resolution}.

\begin{lstlisting}[
    breaklines=true, 
    numbers=left,
    numberstyle = \scriptsize,
    numbersep = 4pt,
    basicstyle = {\normalsize \ttfamily},
    language=XML,
    label=lst:uvl_example,
    frame = lines,
    caption = Example of feature model in UVL format for a specific parser version (mobile phone).]
features
    MOBILE_PHONE {abstract}    
        or
            MP3_Recording
            Camera_Resolution
            
                alternative
                    2.1MP
                    5 MP
                    3.1MP
            Camera
            Audio_Formats 
                or
                    WAV
                    MP3

constraints
    MP3_Recording => MP3
\end{lstlisting}

\begin{figure*}[t]
\begin{center}
	\includegraphics[width = 0.7 \textwidth]{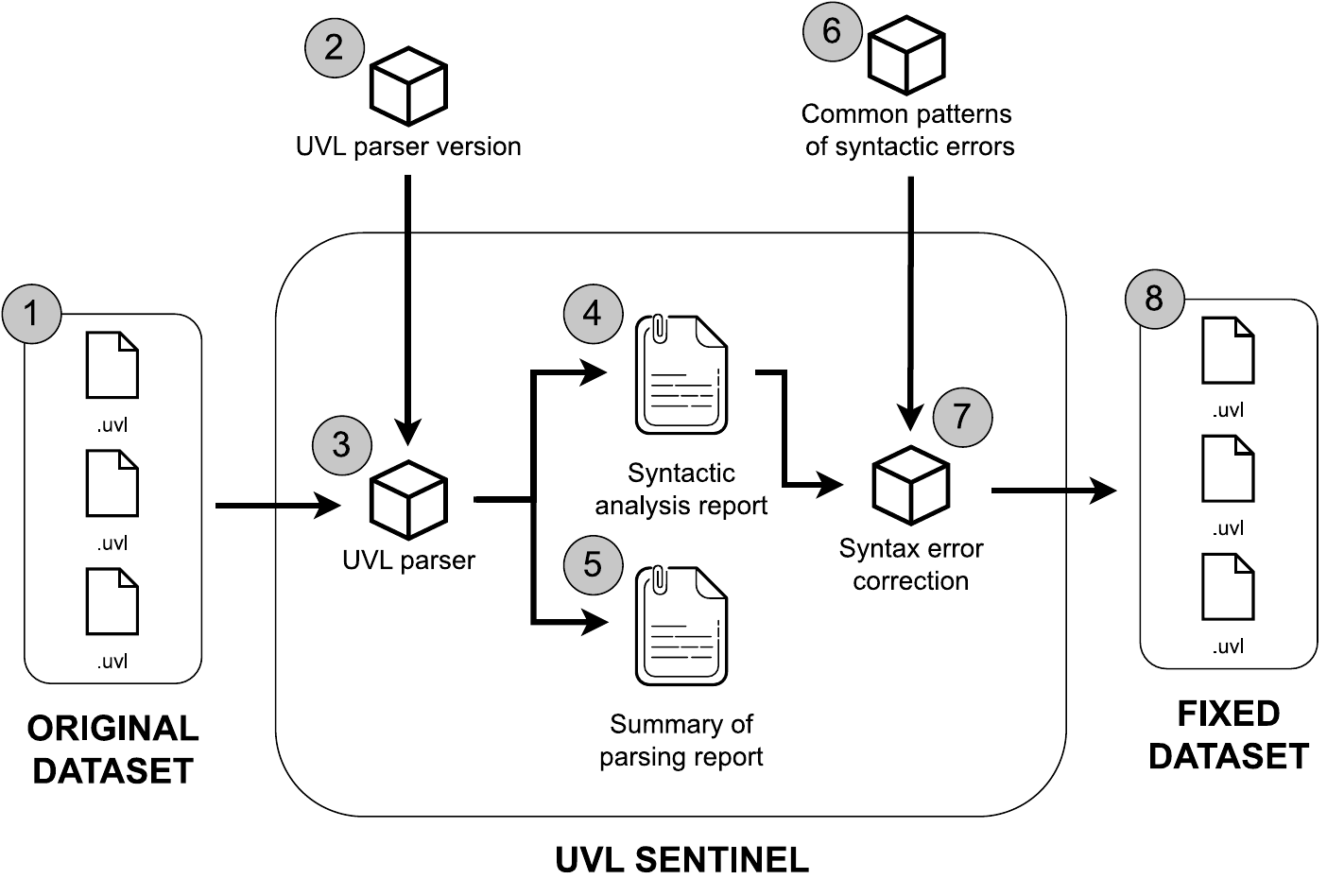}
\end{center}
	\caption{UVL Sentinel main components}
	\label{fig:uvl_syntax_sentinel_components}
\end{figure*}

Within the feature \emph{Camera\_Resolution}, there are three alternatives for the megapixel resolution of the camera: \emph{2.1 MP}, \emph{5 MP} and \emph{3.1 MP} (lines 8-10). However, the feature names contain characters that are not allowed: a dot, ".", a blank space " " , and beginning with a digit instead of a character (5 MP). In addition, a blank line (line 6) is not allowed in the syntax of the current parser version.

Syntactic errors can occur in the UVL transformation process. These errors cause the syntax to be incorrect for the parser. Furthermore, even if the transformation does not cause errors, updating the parser to support new features or syntax changes causes these models to return errors. If a dataset has fewer feature models, less than 20 models, it may be feasible to solve them by hand. However, it is impossible to solve massive datasets with more than 500 models without using some systematic process.

In this paper, we present UVL Sentinel, a tool that analyzes a dataset composed of several feature models in UVL format. This tool generates valuable reports for researchers about possible syntactic errors in the models and helps to fix these errors semi-automatically, thanks to common error patterns.

\section{A tool for syntactic checking of datasets}

Our tool can analyze a dataset with any logical organization structure, generate a report of the previous analysis with a UVL parser and, eventually, try to fix errors by applying regular patterns. The four characteristics in detail are:

\begin{itemize}

    \item \emph{Search for files in the dataset}. Researchers often have a dataset that follows a specific logical structure, such as grouping the feature models according to characteristics, number of valid products, configurations, and subject matter. Therefore, the tool goes through the directory tree in depth until it finds all the files with extension \emph{.uvl}. Each file is an input for the parser, which will dump the parsed information either by console or plain text.
    
    \item \emph{Analysis report generation}. The tool generates a report file with a fundamental analysis of each \emph{.uvl} file found, indicating (1) the dataset to which it belongs, (2) the name of the file with extension \emph{uvl}, (3) whether a warning, an unexpected exception or both have been raised and (4) a description of the error.
    
    \item \emph{Error reporting analysis}. From this text report, the tool counts the number of warnings, exceptions, and errors, among others, found in the report in the previous phase and generates a new report with the error summary.
    
    \item \emph{Correction of common formatting errors}. Many of the syntactic errors and incompatibilities are due to, but not limited to, (1) characters not allowed in the name of features or constraints, (2) blank lines, (3) tabs on a blank line, (4) encoding not allowed in the language. In the case of the UVL language, using tabulators, as in the Python language, is indispensable to distinguish which elements belong to the same block. Tabulators on blank lines, being non-printable characters, are not easy to detect with just a visual glance at the development environment. Our tool detects and corrects these anomalies using regular patterns and performing matching between these patterns and the syntactic errors found.
    
\end{itemize}

\section{Implementation}

Figure \ref{fig:uvl_syntax_sentinel_components} shows the main components of the tool, the inputs, the outputs and the configurations we can make to adapt a dataset to the needs of a specific version of a parser.

\begin{enumerate}

    \item \emph{Original dataset}. We start with a  dataset of an arbitrary number of feature models in UVL format. The organization of these files in directories or subdirectories is unknown to the tool; this organization responds to the logical structure defined by the researchers, and that makes sense within its scope. The tool performs an in-depth search for all files with UVL extension and stores the path for reporting purposes.

    \item \emph{UVL parser version}. If the parser is available through a package manager, the version we want to test can be defined in a dependency file. The tool is abstracted from the version, allowing a report to be generated for different versions of the same parser.

    \item \emph{UVL parser}. After discovering all the UVL files in this component, the tool calls the parser, passing a single UVL file as a parameter. The output, either by console (desirable) or dumped to a file, is processed by the tool and stored in the Syntactic analysis report.

    \item \emph{Syntactic analysis report}. The output of the parser upon receiving each UVL file can generate (1) the plain text representation of the feature model, a symptom that the parser has worked, (2) a warning of possible incompatibilities between the parser version and the current syntax of the UVL file or (3) an exception because it has been impossible to parse the UVL file. The tool generates a report of the full path to each UVL file, whether a warning or exception was raised, and the output generated by the parser to help identify the error.

    \item \emph{Summary of parsing report}. Sometimes, the \emph{Syntactic analysis report} needs to be shorter to be studied at a glance. Therefore, the tool summarizes (1) correctly parsed UVL files, (2) files with warnings and (3) files where an unhandled exception was caught.

    \item \emph{Common patterns of syntactic errors}. Often, many syntactic errors arising from incompatibilities between different versions of a parser are due to disallowed characters, blank lines or problems associated with the file's encoding. For this reason, the tool has a base of regular expressions as error detection patterns. 

    \item \emph{Syntax error correction}. Thanks to the \emph{syntactic analysis report} and the \emph{common patterns of syntactic errors}, the tool helps to match the error obtained when reading a UVL file with the possible solution. For example, the middle hyphen character "-" is not allowed as part of a feature name, and the solution is to replace it with an underscore "\_". The tool processes each UVL individually and generates a new UVL with solutions applied based on the error patterns mentioned in Stage 6.

    \item \emph{Fixed dataset}. The tool starts from the original dataset structure. It makes an exact copy of the files, replacing those original UVL files with the ones generated and fixed in the previous component. This new dataset serves, again, as input for the tool to generate new reports, allowing to compare the number of errors before using the tool and after and to study if there are significant differences.
    
\end{enumerate}

\section{Validation}

For the validation of the tool, we used 1,479 feature models in UVL format organized in 20 datasets. We have performed a massive analysis of all models, generated reports before and after the syntactic correction and analyzed them.

\subsection{Warning, errors and exceptions}

For a correct understanding of the validation, the following terms are defined:

\begin{itemize}

    \item \emph{Warning}. A warning is a warning from the parser that there are inconsistencies in the syntax but that it is possible to build the model.

    \item \emph{Error}. An error is a message that there is a syntactically incorrect element and must be corrected. 

    \item \emph{Exception}. When an exception occurs, the parser cannot read the model because of a formatting problem.
    
\end{itemize}

\subsection{Analysis of the dataset before syntax error correction}

The data from the syntactic analysis report before running the syntactic correction have been:

\begin{itemize}

    \item \emph{Syntactic analysis report}. We have analyzed the 1,479 and found descriptive syntactic errors such as (1) invalid characters (\emph{extraneous input}), (2) blank lines (\emph{mismatched input}), and (3) characters not allowed as beginnings of feature names (\emph{token recognition}).

    \item \emph{Summary of parsing report}. The total number of UVL files with warnings is 0 (0.00\% of total files), and the total number of UVL files with exceptions is 185 (12.51\%).
    
\end{itemize}

\subsection{Analysis of the dataset after syntax error correction}

The data from the syntactic analysis report after running the syntactic correction have been:

\begin{itemize}

    \item \emph{Syntactic analysis report}. Visually, the full errors report reflects that there has been an improvement in the number of syntactic problems. However, we still see errors such as \emph{extraneous input} in some UVL files.

    \item \emph{Summary of parsing report}.  The total number of UVL files with warnings is 0 (0.00\% of total files), and the total number of UVL files with exceptions is 25 (1.69\%).
    
\end{itemize}

We note that UVL Sentinel has helped to detect and correct several syntactic errors by 86.49\%. However, there are errors that we have not contemplated in the set of defined patterns because they are more difficult to define or do not contemplate all the possible options of syntactic variants. Our tool is limited to defining this set of patterns and their associated solutions. A detailed study covering more syntactic error detection patterns is needed.

\section{Conclusion and future work}

UVL Sentinel allows a comprehensive analysis of datasets with an arbitrary structure of feature models in UVL format, generates valuable reports for researchers and assists in the semi-automatic syntactic update of the model set of one or several datasets to newer versions of a parser.

Our tool prevents a valuable dataset for the scientific community from being tied to a particular parser version. It can no longer be used as a study element because it is incompatible with new UVL parsers or the same parser but with an updated version that generates incompatibilities in its syntax. However, the success of syntax correction is limited to the set of regular expression definitions and their possible solutions. A systematic and in-depth study of all syntactic errors with a large volume of UVL models is needed to determine the best strategy for correcting datasets and to keep them as part of future studies on feature models and software product lines.

In future work, the idea is proposed to be extended to other UVL parsers or feature model description languages with syntactic incompatibility problems. It would also be helpful to have more error detection patterns and more detailed reports, such as parse time or a cloud of the most common errors encountered when reading models from a dataset.

\section*{Material}

The developed software, example datasets and deployment instructions can be obtained from this repository: \newline https://github.com/diverso-lab/uvl-sentinel

\section*{Acknowledgments}

This work was partially supported by FEDER/Ministry of Science, Innovation and Universities/Junta de Andalucía/State Research Agency/CDTI with the following grants: \emph{Data-pl} (PID2022-138486OB-I00), TASOVA PLUS research network (RED2022-134337-T) and 
MIDAS (IDI-20230256)

\bibliographystyle{ACM-Reference-Format}
\bibliography{references}

\end{document}

%% file: __packages__.tex
\usepackage{url}
\usepackage{hyperref}

\usepackage{enumitem} 

\usepackage{dblfloatfix} 

\usepackage{threeparttable}
\usepackage{booktabs}
\usepackage{makecell}
\usepackage{multirow}
\usepackage{tabularx}
\usepackage{pgfplotstable}
\usepackage{xcolor}
\usepackage{siunitx}
\usepackage{colortbl}

\usepackage{listings}
\usepackage{algorithm}
\usepackage{algpseudocode}

\usepackage{marvosym} 
\usepackage{bm} 
\usepackage{dsfont}

\newcommand{\NA}[1]{NA}

\usepackage{setspace}
\usepackage{adjustbox}

\usepackage{amsthm}
\usepackage{thmtools}
\declaretheoremstyle[
    notefont=\bfseries,
    qed=$\blacksquare$
]{mydefistyle}